\documentclass[11pt,onecolumn]{article} 
\usepackage{latex8}
\usepackage{times}

\usepackage{algorithm}
\usepackage{algorithmic}
\usepackage{graphicx}
\usepackage{amsmath}
\usepackage{amssymb}

\begin{document}

\title{Computing Multiplicative Order and Primitive Root in Finite Cyclic Group}


\author{
 Shri Prakash Dwivedi \thanks{\texttt{Email: shriprakashdwivedi@gbpuat-tech.ac.in}}\\ 
}

\maketitle

\begin{abstract}
Multiplicative order of an element $a$ of group $G$ is the least positive integer $n$ such that $a^n=e$, where $e$ is the identity element of $G$. If the order of an element is equal to $|G|$, it is called generator or primitive root. This paper describes the algorithms for computing multiplicative order and primitive root in $\mathbb{Z}^*_{p}$, we also present a logarithmic improvement over classical algorithms.
\end{abstract}

\section{Introduction}
Algorithms for computing multiplicative order or simply order and primitive root or generator are important in the area of random number generation and discrete logarithm problem among others. In this paper we consider the problem of computing the order of an element over $\mathbb{Z}^*_{p}$, which is multiplicative group modulo $p$ . As stated above order of an element $a \in \mathbb{Z}^*_{p}$ is the least positive integer $n$ such that $a^n=1$. In number theoretic language, we say that the order of $a$ modulo $m$ is $n$, if $n$ is the smallest positive integer such that $a^n \equiv 1(\mod  m)$. We also consider the related problem of computing the primitive root in $\mathbb{Z}^*_{p}$. If order of an element $a \in \mathbb{Z}^*_{n}$ usually denoted as $\text{ord}_n(a)$ is equal to $|\mathbb{Z}^*_{n}|$ i.e. order of multiplicative group modulo $n$, then $a$ is called called primitive root or primitive element or generator [3] of $\mathbb{Z}^*_{n}$. It is called generator because every element in $\mathbb{Z}^*_
{n}$ is some power of $a$.\\
\indent Efficient deterministic algorithms for both of the above problems are not known yet. However if the prime factorization of $\phi(n)=|\mathbb{Z}^*_{n}|$ is provided then efficient algorithms can be designed. Since factorization itself is very difficult for large numbers, and no polynomial time algorithm is known for this problem. Therefore no direct method is available to solve these problems when the size of the group or $n$ is very large.\\
\indent Work has been done on searching for primitive root in $\mathbb{F}_{p^n}$. Here the task is to generate a subset of $\mathbb{F}_{p^n}$, which contains at least one primitive root [8, 9]. Assuming Extended Riemann Hypothesis (ERH) it has been shown that there exists a positive integer $n=(\log p)^c$ for some constant $c$ such that $n\mod p$ is primitive root over $\mathbb{F}_{p}$ [10]. However searching for small primitive root not necessarily imply a fast method for computing primitive root. In [5] authors presented a randomized algorithm for generating primitive root modulo a prime with high probability, in particular the algorithm computes every prime factor $p_i$ of $p-1$ such that $p_i$ is less than some specified value.\\
\indent For computing order and primitive root in $\mathbb{Z}^*_{p}$, factorization of order of the group $|\mathbb{Z}^*_{p}|$ is required, and as we mentioned before that factorization of $\phi(p)=|\mathbb{Z}^*_{p}|=p-1$ can not be calculated efficiently for large $p$, and there is no any other approach to attack the problem, it has been suggested to construct or generate a large prime $p$ together with primitive root for $\mathbb{Z}^*_{p}$. In this setting prime factorization of $(p-1)$ is known and the task is to compute primitive root with high probability. This paper describes the algorithms in this context. We describe straight forward basic algorithms as well as a logarithmic improvement over the traditional one.\\
\indent This paper is organized as follows. Section II explains preliminaries and basic algorithms, section III describes modified algorithms and their analysis, finally, section IV contains conclusion.

\section{Preliminaries}
A group $(G, *)$ is an algebraic structure, which consists of a set $G$ together with a binary operation * over $G$, such that * follows closure, associative property, $G$ possesses a unique identity element $e$, and every element $a$ of $G$ has unique inverse $a^{-1}$. When the binary operation * is clear from context, the group is simply represented by $G$. Order or size of a group is the number of elements in $G$ and denoted as $o(G)=|G|$. If order of a group is a finite number, then it is called finite group. If $G$ is a group then order of $a \in G$ is the least positive integer $n$ such that $a^n=e$.\\
\indent The set $\mathbb{Z}_n=\{0,1,...,n-1\}$ under addition modulo $n$ forms a group where equivalence class $[0]_n$ is the identity and equivalence class $[-a]_n$ is the inverse of $[a]_n$. The set $\mathbb{Z}_n^* = \{ a \in \mathbb{Z}_n | gcd(a,n)=1 \}$ or $\mathbb{Z}_n^* = \{ 0<a<n | gcd(a,n)=1 \}$ under multiplication modulo $n$ forms a group with equivalence class $[1]_n$ as identity, and inverse of $[a]_n$ is denoted by $[a]_n^{-1}$.\\
\indent A multiplicative group $G$ is said to be cyclic group if $G=\langle a \rangle=\{ a^n | n \in \mathbb{Z} \}$, it implies that there exists $a \in G$ such that for every $b \in G$ there exists $n$ with $b=a^n$. Here $a$ is called generator or primitive root or primitive element. By definition every cyclic group consists of a generator. For example additive group $\mathbb{Z}_n$ is finite cyclic group with equivalence class $[1]_n$ as a generator. Now we state following results, which can be found in any standard algebra texts [4, 7].\\

\textbf{Proposition 1}. \textit{Let $G$ be a finite group and $H$ is a subgroup of $G$, then $o(H)|o(G)$}.\\ 
\indent Let $H = \{a_1,a_2,...,a_n\}$, here $|G|=n$. Let there be an element $b \in G$ and $b \notin H$, now by taking product of $b$ and elements of subgroup $H$, we can create $n$ new and distinct elements of $G$. Which are $\{ba_1, ba_2,...,ba_n\}$. Note that if $ba_i=ba_j$, it imply that $a_i=a_j$, which is not possible since all $a_i$'s are distinct by definition. Also $ba_i=a_j \Rightarrow b=a_ja_i^{-1}$, and since $H$ is subgroup, therefore by definition it is closed, and every element has a inverse. It imply that $a_ja_i^{-1} \in H$ again a contradiction. By repeating in this way for every new element of $G$ which is not already in $H$, we can produce $n$ more new and distinct elements of $G$. Suppose we stop after $m$ iterations then $|G|=m|H|$ and therefore $|H|$ divides $|G|$.\\

\textbf{Proposition 2}. \textit{Let $G$ be a finite group and $a \in G$, then $o(a)|o(G)$}.\\ 
\indent Since $\langle a \rangle = \{a^n | n \in \mathbb{Z}\}$ is a subgroup which happened to be cyclic and generated by $a$. $|a|$ divides $|G|$.\\

\textbf{Proposition 3}. \textit{Let $G$ be a finite group and $a \in G$, then $a^{o(G)}=e$}.\\
\indent Using Proposition 2, we can write $|G| = m |a|$ for $m \in \mathbb{Z}^+$. Hence $a^{|G|}=a^{m.|a|}=(a^{|a|})^m=e^m=e$.
Above Proposition in number-theoretic context can be stated as follows.\\

\textbf{Proposition 4}. (Euler's Theorem) \textit{If $a$ is relatively prime to a positive integer $n$, then $a^{\phi(n)} \equiv 1(\mod n)$ for all $a \in \mathbb{Z}_n^*$}.\\ 
\indent Since $\mathbb{Z}_n^*$ is multiplicative group with $|\mathbb{Z}_n^*|=\phi(n)$ and identity 1.\\

\textbf{Proposition 5}. (Fermat's Theorem) \textit{$a^{p} \equiv a(\mod p)$ for any prime $p$ and all $a \in \mathbb{Z}_p^*$}.\\ 
\indent Restricting $n$ to prime number $p$ and putting $\phi(p)=p-1$ in Euler's theorem, Proposition 5 follows.\\

\textbf{Proposition 6}. \textit{Let $G$ be a finite group whose order is a prime number then $G$ is a cyclic group}.\\ 
\indent Here $|G|$ is a prime number. Suppose $a \in G$ and is distinct from $e$. From Proposition 2 $o(\langle a \rangle)|o(G)$ and $o(\langle a \rangle) \neq 1$. It follows that $|\langle a \rangle|=|G|$. \\

\textbf{Proposition 7}. \textit{The multiplicative group $\mathbb{Z}_n^*$ is cyclic, if $n$ equals to 2, 4, $p^e$ and $2p^e$ for any odd prime $p$ and positive integer $e$} [6]. \\ 

\textbf{Proposition 8}. \textit{Let $a,b \in \mathbb{Z}_n^*$ such that $a$ has order $n_1$, $b$ has order $n_2$ and $gcd(n_1,n_2)=1$, i.e. $n_1$ and $n_2$ are relatively prime, then $a.b$ has order $n_1.n_2$}. \\ 
\indent We have $(ab)^{n_1n_2}=a^{n_1n_2}b^{n_1n_2}=(a^{n_1})^{n2}(b^{n_2})^{n1}=1$. Therefore $o(ab)|n_1n_2$. Let $m=o(ab)$, then $b^{n_1m}=(a^{n_1})^{m}(b^{n_1})^{m}=((ab)^m)^{n_1}=1 $. Hence $n_2|n_1m$ but $gcd(n_1,n_2)=1$ so $n_2|m$. Similarly $n_1|m$ and therefore $n_1n_2|m$.

\textbf{Proposition 9}. \textit{Let $a \in \mathbb{Z}_n^*$.  If $a^{p^e}=1$ and $a^{p^{e-1}} \neq 1$ for some prime $p$ and $e \in \mathbb{Z}^+$, then $a$ has order $p^e$}. \\ 
\indent Let $m$ be the multiplicative order of $a$, that is $m$ is the least positive integer such that $a^m=1$. If $a^{p^e}=1$ then, $m | p^e$. Since $p$ is prime, let $p^e = m.p^{e'}$ then $e'$ should be one of $0,1,2,...,e$. In the case $e'<e$, it imply that $a^{p^{e-1}} = 1$, which is contradiction and therefore $e'=e$.\\

\textbf{Proposition 10}. \textit{Let $a \in \mathbb{Z}_p^*$ and $a^{p-1}=1$. Let prime factorization of $p-1$ be $p_1^{e_1}p_2^{e_2}...p_k^{e_k}$. Let $m_i$ be the largest integer such that $a^{(p-1)/p_i^{m_i}}=1$, then order of $a$ is $p_1^{e_1-m_1}p_2^{e_2-m_2}...p_k^{e_k-m_k}$} [1]. \\

Algorithms for computing order and primitive root can be found in any standard computational number theory and related books [1, 2, 7]. In this section we describe straight forward algorithms to perform these tasks.
Computation of Multiplicative-Order is described in Algorithm 1. Input to this algorithm are prime factorization of order of finite cyclic group $|\mathbb{Z}_p^*|=p-1=p_1^{e_1}p_2^{e_2}...p_k^{e_k}$, along with an element $a$ of this group $\mathbb{Z}_p^*$. Output to this algorithm is multiplicative order of $a$.

\begin{algorithm}
\caption{\bf :  Multiplicative-Order $(\mathbb{Z}_p^*, a)$}
\begin{algorithmic}
\STATE \textbf{INPUT}: $|\mathbb{Z}_p^*|=p-1=p_1^{e_1}p_2^{e_2}...p_k^{e_k}, a \in \mathbb{Z}_p^*$
\STATE \textbf{OUTPUT}: Multiplicative order $n$ of $a$
   \STATE $n \leftarrow p-1 $ \\
   \FOR {$(i\leftarrow 1; i \leq k; i\leftarrow i+1)$ }
   {
   \STATE $n \leftarrow (p-1)/p_i^{e_i} $ \\
   \STATE $b \leftarrow a^n $ \\
   \WHILE { $(b \neq 1)$ }
   {
   \STATE $b \leftarrow b^{p_i} $ \\
   \STATE $n \leftarrow n*p_i $ \\
   }
   \ENDWHILE
   }
   \ENDFOR
   \RETURN $n$

\end{algorithmic}
\end{algorithm}

Algorithm 2 describes Primitive-Root computation. Input to this algorithm is prime factorization of order of group $\mathbb{Z}_p^*$, and output to this algorithm is primitive root of this group. Primitive-Root is a randomized algorithm as it selects a random element $a$ of $\mathbb{Z}_p^*$ in the first step of each iteration.

\begin{algorithm}
\caption{\bf :  Primitive-Root $(\mathbb{Z}_p^*)$}
\begin{algorithmic}
\STATE \textbf{INPUT}: $|\mathbb{Z}_p^*|=p-1=p_1^{e_1}p_2^{e_2}...p_k^{e_k}*$
\STATE \textbf{OUTPUT}: Primitive root $a$ of $\mathbb{Z}_p^*$
   \STATE Select $a \in \mathbb{Z}_p^*$ at random \\
   \FOR {$(i\leftarrow 1; i \leq k; i\leftarrow i+1)$ }
   {
   \STATE $b \leftarrow a^{(p-1)/p_i} $ \\
   \IF {$(b==1)$}
   \STATE Primitive-Root $(\mathbb{Z}_p^*)$
   \ENDIF
   }
   \ENDFOR
   \RETURN $a$

\end{algorithmic}
\end{algorithm}

\section{Algorithms}
\subsection{Computing Multiplicative Order}
For computing multiplicative order of an element $a \in \mathbb{Z}_n^*$, where prime factorization of $n$ is given as $$n=n_1*n_2*...*n_k= \prod_{i} n_i$$ and we are required to compute $(a^{n/n_1}, a^{n/n_2},..., a^{n/n_k})$. Let $n_i'=n/n_i$ for $i=1,...,k$. Therefore $(a^{n/n_1}, a^{n/n_2},..., a^{n/n_k})= (a^{n_1'},a^{n_2'},...,a^{n_k'})$. Here we assume that $n_i'$ is calculated as $n_i'=n/n_i=n_1*n_2*...*n_{i-1}*n_{i+1}*...*n_k$. To compute $n_i'$, $k-2$ multiplications are required. For example, to compute $n_1'=n/n_1=n_2*n_3*...*n_k$, it requires $k-2$ multiplications. By using some precomputations $n_i'$ can be computed in only $\log k$ multiplications. Therefore total cost to compute $a^{n_i'}$ becomes $O(\log k.(\log n)^3)$ bit operations.\\
\indent For $k=4$, we have $n=n_1*n_2*n_3*n_4$. With precomputing:\\
$N_{12}=n_1*n_2$\\
$N_{34}=n_3*n_4$\\
We can compute each $n_i'$ in only two multiplications.\\
$n_1'=n_2*N_{34}$\\
$n_2'=n_1*N_{34}$\\
$n_3'=N_{12}*n_4$\\
$n_4'=N_{12}*n_3$\\ \\
Similarly for $k=8$, we have \\
$N_{12}=n_1*n_2$, $N_{34}=n_3*n_4$\\
$N_{56}=n_5*n_6$, $N_{78}=n_7*n_8$\\
$N_{1234}=N_{12}*N_{34}$, $N_{5678}=N_{56}*N_{78}$\\
Now by using above precomputations, we can compute each $n_i'$ is only $\log 8 -1=2$ multiplications.\\
$n_1'=n_2*N_{34}*N_{5678}$\\
$n_2'=n_1*N_{34}*N_{5678}$\\
$n_3'=N_{12}*n_4*N_{5678}$\\
$n_4'=N_{12}*n_3*N_{5678}$\\
$n_5'=N_{1234}*n_6*N_{78}$\\
$n_6'=N_{1234}*n_5*N_{78}$\\
$n_7'=N_{1234}*N_{56}*n_8$\\
$n_7'=N_{1234}*N_{56}*n_7$\\
Above method is generalized in the Algorithm 3. Input to K-Exponentiation algorithm is $n \in \mathbb{Z}^+$ along with with its $k$ factors. Here, we assume that $k$ is exact power of some positive integer that is $k=\{2^m|m \in \mathbb{Z}^+\}$.  Output of this algorithm is $k$ integers $a_{(1...k)}$ such that $a_i=a^{n_i'}$, where $n_i'=n/n_i=n_1*n_2*...*n_{i-1}*n_{i+1}*...*n_k$. Brief description of this algorithm is as follows. First precomputed values are assigned in $N_{12}, N_{1234} \text{ to } N_{12...k/2}$ etc. For loop is used to compute $n_i'$ values for $i=1,2,...,k$, First If loop is used to check whether $i \leq k/2$ depending on that second (inner) If loop is used to check whether $i$ is odd or even. If $i$ is odd $n_i'$ is calculated in If loop, otherwise it is calculated in Else loop. Again this calculation is repeated where $i > k/2$ in Else (outer) loop.

\begin{algorithm}
\caption{\bf :  K-Exponentiation $(n, a)$}
\begin{algorithmic}
\STATE \textbf{INPUT}: $n=n_1*n_2*...*n_k, a \in \mathbb{Z}_n^*$
\STATE \textbf{OUTPUT}: $a_{(1...k)}=(a^{n_1'},a^{n_2'},...,a^{n_k'})$
  \STATE $N_{12} \leftarrow n_1*n_2, N_{34} \leftarrow n_3*n_4,...,N_{(k-1)k} \leftarrow n_{(k-1)}*n_k$\\
  \STATE $N_{1234} \leftarrow N_{12}*N_{34},..., N_{(k-3)(k-2)(k-1)k} \leftarrow N_{(k-3)(k-2)}*N_{(k-1)k}$ \\
  \STATE ..........
  \STATE Compute $N_{12...k/2}$, $N_{(k/2+1)(k/2+2)...k}$
   \FOR {$(i\leftarrow 1; i \leq k; i\leftarrow i+1)$ }
   {
   \IF {$(i \leq k/2)$}
   \IF {$(i \mod 2==1)$}
   \STATE $n_i'=N_{12...k/4}...N_{(i-2)(i-1)}*$ \\
   \STATE      $ N_{i+1}*(N_{(i+2)(i+3)})...N_{(k/2+1)...k}$ \\
   \ELSE 
   \STATE $n_i'=N_{12...k/4}...N_{(i-3)(i-2)}*$ \\
   \STATE      $ N_{i-1}*(N_{(i+1)(i+2)})...N_{(k/2+1)...k}$ \\
   \ENDIF
   \ELSE 
   \IF {$(i \mod 2==1)$}
   \STATE $n_i'=N_{12...k/2}...N_{(i-2)(i-1)}*$ \\
   \STATE      $ N_{i+1}*(N_{(i+2)(i+3)})...N_{(3k/4+1)...k}$ \\
   \ELSE 
   \STATE $n_i'=N_{12...k/2}...N_{(i-3)(i-2)}*$ \\
   \STATE      $ N_{i-1}*(N_{(i+1)(i+2)})...N_{(3k/4+1)...k}$ \\
   \ENDIF
   \ENDIF
   }
   \ENDFOR
   \STATE Compute $a_{(1...k)}=(a^{n_1'},a^{n_2'},...,a^{n_k'})$
   \RETURN $a_{(1...k)}$

\end{algorithmic}
\end{algorithm}

Correctness of K-Exponentiation algorithm can be easily established using induction on number of products $k$. \\

\textbf{Theorem 1.} \textit{K-Exponentiation algorithm computes $a_i=a^{n_i'}$ for $i=1,2,...,k$ where $n_i'=n_1*n_2*...*n_{i-1}*n_{i+1}*...*n_k$. } \\

\textit{Proof}: Assume $k=\{2^m|m \in \mathbb{Z}^+\}$. Since $2^m$ is the number of products in our case, we shall use induction on $m$. For the base case we take $m=1$, therefore we have $k=2^m=2$. It is a trivial case. Here $n_1'=n_2$ and $n_2'=n_1$. As a induction hypothesis assume that the above statement is true for upto $m=r$. Given the statement for $k=2^r$. We can construct the products for $k=2^{r+1}$. Note that in case of $k=2^r$, we have two products of length $2^{r/2}$ which are $n_1*n_2*...*n_{2^{r/2}}$ and $n_{2^{(r/2)+1}}*...*n_{2^{r}}$. Using these products we can construct $n_1*n_2*...*n_{2^r}= (n_1*n_2*...*n_{2^{r/2}})*(n_{2^{n/2+1}}*...*n_{2^{r}})$. It is the first product of length $2^r$. For the second product, we need two more construction of length $2^{r/2}$. It is $n_{(2^r)+1}*n_{(2^r)+2}*...*n_{(2^r)+(2^r)}=(n_{(2^r)+1}*n_{(2^r)+2}*...*n_{(2^r)+(2^r)/2})*(n_{(2^r)+((2^r)/2)+1}*...*n_{(2^r)+(2^r)})$. Now we have constructed both products of length $2^r$ using the products of length 
$2^r/2=2^r-1$, and the statement follows for $k=2^{r+1}$. Now, using $n_i'$, we can compute $a_i=a^{n_i'}$ for $i=1,2,...,k$.\\

\indent Complexity to compute $a^{n_i'}$ for $a \in \mathbb{Z}_n^*$ is $O(\log k.(\log n)^3)$ operations. Note that, we can compute $n_i'$ in $O(\log k)$ operations. whereas $a^{n_i'}$ can be computed using repeated squaring algorithm for modular exponentiation in $O((\log n)^3)$ operations for $a \in \mathbb{Z}_n^*$. In general Algorithm 3 performs $(\log k -1).\log k$ precomputations and using that it calculates each $n_i'$ in $(\log k -1)$ multiplications.\\
\indent Using K-Exponentiation algorithm as a subroutine, we can write the Modified-Multiplicative-Order algorithm. It is described in Algorithm 4. Again, input to this algorithm are prime factorization of order of group $|\mathbb{Z}_p^*|=p-1=p_1^{e_1}p_2^{e_2}...p_k^{e_k}$, along-with an element $a \in \mathbb{Z}_p^*$. Output to this algorithm is multiplicative order of $a$. First step of this algorithm calls K-Exponentiation to compute $a_i=a^{p_i^{e_i}}$ for $i=1,...,k$ and stores it in list $a_{(1...k)}$. In the second step $m_i$ is initialized to 0, for $i=1,...,k$. After that for each $i$ in the while loop maximum integer $m_i$ is calculated such that $a^{(p-1)/p_i^{m_i}}=1$, and using that final order is computed.\\

\begin{algorithm}
\caption{\bf :  Modified-Multiplicative-Order $(\mathbb{Z}_p^*, a)$}
\begin{algorithmic}
\STATE \textbf{INPUT}: $|\mathbb{Z}_p^*|=p-1=p_1^{e_1}p_2^{e_2}...p_k^{e_k}, a \in \mathbb{Z}_p^*$
\STATE \textbf{OUTPUT}: Multiplicative order $n$ of $a$
   \STATE $a_{(1...k)} \leftarrow $ K-Exponentiation $(p_1^{e_1}p_2^{e_2}...p_k^{e_k}, a)$\\
   \STATE $m_{(1...k)} \leftarrow 0_{(1...k)} $ \\
   \FOR {$(i\leftarrow 1; i \leq k; i\leftarrow i+1)$ }
   {
   \WHILE { $(a_i^{p_i} \neq 1)$ }
   {
   \STATE $a_i \leftarrow a_i^{p_i} $ \\
   \STATE $m_i \leftarrow m_i +1 $ \\
   }
   \ENDWHILE
   }
   \ENDFOR
   \RETURN $n=p_1^{e_1-m_1}p_2^{e_2-m_2}...p_k^{e_k-m_k}$

\end{algorithmic}
\end{algorithm}

\textbf{Theorem 2.} \textit{Algorithm 4 computes multiplicative order of $a \in \mathbb{Z}_p^*$.} 

\textit{Proof}: Statement of the theorem follows from Proposition 8 and 10.\\
\indent Overall complexity of this algorithm is dominated by computing $a_i=a^{p_i^{e_i}}$, which is $O(\log k.(\log p)^3)$ bit operations in $\mathbb{Z}_p^*$.

\subsection{Computing Primitive Root}

Primitive root of a finite cyclic group is an element whose order is equal to size of the group. From this basic definition of primitive root itself, we can write a simple algorithm, which select a random element $a \in \mathbb{Z}_p^*$ and computes it's multiplicative order. If multiplicative order is equal to $\phi(p)=p-1$ then it is one of the primitive root. This method is summarized in Algorithm 5 named as Simple-Primitive-Root. In Algorithm 5, If loop uses to check whether order of $a$ is equal to $p-1$, if it is the case $a$ is returned otherwise algorithm calls itself and go to first step.  If we want to find out least primitive root, then instead of choosing an element randomly, better way is to start from least value of $a$ to consecutive higher value $a+1, a+2,...$ etc. \\

\begin{algorithm}
\caption{\bf :  Simple-Primitive-Root $(\mathbb{Z}_p^*)$}
\begin{algorithmic}
\STATE \textbf{INPUT}: $|\mathbb{Z}_p^*|=p-1=p_1^{e_1}p_2^{e_2}...p_k^{e_k}*$
\STATE \textbf{OUTPUT}: Primitive root $a$ of $\mathbb{Z}_p^*$
   \STATE Select $a \in \mathbb{Z}_p^*$ at random \\
   \STATE $m \leftarrow $ Multiplicative-Order $(\mathbb{Z}_p^*, a)$\\
   \IF {$(m==p-1)$}
   \RETURN $a$
   \ELSE
   \STATE Simple-Primitive-Root $(\mathbb{Z}_p^*)$
   \ENDIF

\end{algorithmic}
\end{algorithm}

\indent While the above algorithm for computing primitive root using multiplicative order is simple, other methods are also there to find primitive element. One such method we have seen in Algorithm 2. Now we describe the Modified-Primitive-Root algorithm using K-Exponentiation. It is outlined in Algorithm 6. Modified-Primitive-Root is almost same as Algorithm 2, except that it calls K-Exponentiation to compute $a_i=a^{p_i^{e_i}}$ for $i=1,...,k$ and stores it in list $a_{(1...k)}$. At any time in the If loop, whenever algorithm detects $a_i=1$, it calls itself and go to step 1 and chooses another random element.\\

\begin{algorithm}
\caption{\bf :  Modified-Primitive-Root $(\mathbb{Z}_p^*)$}
\begin{algorithmic}
\STATE \textbf{INPUT}: $|\mathbb{Z}_p^*|=p-1=p_1^{e_1}p_2^{e_2}...p_k^{e_k}*$
\STATE \textbf{OUTPUT}: Primitive root $a$ of $\mathbb{Z}_p^*$
   \STATE Select $a \in \mathbb{Z}_p^*$ at random \\
   \STATE $a_{(1...k)} \leftarrow $ K-Exponentiation $(p_1^{e_1}p_2^{e_2}...p_k^{e_k}, a)$\\
   \FOR {$(i\leftarrow 1; i \leq k; i\leftarrow i+1)$ }
   {
   \IF {$(a_i==1)$}
   \STATE Modified-Primitive-Root $(\mathbb{Z}_p^*)$
   \ENDIF
   }
   \ENDFOR
   \RETURN $a$

\end{algorithmic}
\end{algorithm}

\indent Correctness of Algorithm 6 follows from Proposition 2, 8 and 9. Again the algorithm is dominated by the computation of $a_i=a^{p_i^{e_i}}$, which using K-Exponentiation is $O(\log k.(\log p)^3)$ operations instead of $O(k.(\log p)^3)$ operations. These randomized algorithms works particularly because for a prime $p$, $\mathbb{Z}_p^*$ has $\phi(\phi(p))=\phi(p-1)$ primitive roots.


\section{Conclusion}
This paper described the algorithms for computing multiplicative order and primitive root in finite cyclic group. It also presented K-Exponentiation algorithm as a subroutine to compute order and primitive elements. In general if the prime factorization of $\phi(p)=p-1$ is given, or $\mathbb{Z}_p^*$ is constructed in such a way that factors of $p-1$ is available, then efficient algorithms can be designed to compute order and primitive roots.

%
%

\end{document}